\documentclass[final,onecolumn,3p,times,sort&compress,12pt]{elsarticle}

\usepackage{graphicx}
\usepackage{dcolumn}
\usepackage{bm}
\usepackage{epstopdf}
\usepackage[breaklinks=true]{hyperref}
\usepackage{bm,color}

\usepackage{doi}

\usepackage{amssymb}
\usepackage{hyperref}
\usepackage{epstopdf,xcolor}

	\usepackage{fancyheadings}
\renewcommand{\thispagestyle}[1]{} 

\begin{document}

\begin{frontmatter}


\title{Thermodynamics of a model solid with magnetoelastic coupling}


\author[a1]{K. Sza\l{}owski\corref{cor1}}
\ead{kszalowski@uni.lodz.pl}
\author[a1]{T. Balcerzak}
\ead{t\_balcerzak@uni.lodz.pl}
\author[a3]{M. Ja\v{s}\v{c}ur}
\address[a1]{Department of Solid State Physics, Faculty of Physics and Applied Informatics,\\
University of \L\'{o}d\'{z}, ulica Pomorska 149/153, 90-236 \L\'{o}d\'{z}, Poland}

\address[a3]{Department of Theoretical Physics and Astrophysics, Faculty of Science, \\P. J. \v{S}\'{a}f\'{a}rik University, Park Angelinum 9, 041 54 Ko\v{s}ice, Slovak Republic}

\cortext[cor1]{Corresponding author}


\date{\today}

\begin{abstract}
In the paper a study of a model magnetoelastic solid system is presented. The system of interest is a mean-field magnet with nearest-neighbour ferromagnetic interactions and the underlying s.c. crystalline lattice with the long-range Morse interatomic potential and the anharmonic Debye model for the lattice vibrations. The influence of the external magnetic field on the thermodynamics is investigated, with special emphasis put on the consequences of the magnetoelastic coupling, introduced by the power-law distance dependence of the magnetic exchange integral. Within the fully self-consistent, Gibbs energy-based formalism such thermodynamic quantities as the entropy, the specific heat as well as the lattice and magnetic response functions are calculated and discussed. To complete the picture, the magnetocaloric effect is characterized by analysis of the isothermal entropy change and the adiabatic temperature change in the presence of the external pressure.
\end{abstract}

\begin{keyword}
magnetoelastic coupling \sep ferromagnetism \sep thermodynamics of magnets \sep magnetocaloric effect \sep entropy \sep specific heat \sep thermodynamic response functions \sep magnetostriction
\end{keyword}


\end{frontmatter}


\newpage
\section{Introduction}

The interplay between the magnetic properties and the elastic and structural characteristics of the underlying lattice in magnetic systems has attracted considerable attention, including both the theoretical and the experimental approach. The influence of external magnetic field on the lattice characteristics as well as the effect of the external pressure on the purely magnetic quantities are the manifestations of the magnetoelastic coupling. The phenomena emerging due to the mutual relations of magnetism and lattice properties were studied in numerous, general theoretical works  \cite{bean_magnetic_1962,callen_static_1963,callen_magnetostriction_1965,brown_theory_1965,alben_magnetoelastic_1969,boubcheur_effects_2001,del_moral_magnetostriction_2007,lu_general_2015,balcerzak_self-consistent_2017,buchelnikov_magnetoelastic_2002,da_silva_effects_2015,plaza_exchange_2007}. Moreover, some special attention was paid to model ferromagnetic materials, such as gadolinium, which have shown further a high potential for applications in magnetocaloric refrigeration. For this substance, both anomalies in thermal expansion \cite{birss_thermal_1960,wakiyama_anomalous_1962,robinson_critical_1978,hadrich_anomalies_1985,bodriakov_magnetoelastic_1992,tokuo_anomalous_2013,kozlovskii_thermal_2015} and in compressibility \cite{freyne_magnetoelastic_1972,klimker_effect_1973,bodryakov_field_1998,spichkin_elastic_1999} were measured carefully, together with the characterization of the phenomenon of magnetostriction \cite{m._bozorth_magnetostriction_1963,coleman_forced_1965,mishima_anomalous_1976}. In addition, the thorough studies of thermodynamic quantities such as entropy \cite{dankov_magnetic_1996}, specific heat \cite{simons_specific_1974,salamon_scaling_1977,bednarz_heat_1993} or magnetic susceptibility \cite{geldart_anisotropy_1989,dunlap_critical_1994} were conducted. This brief review of data proves the importance of searching for models describing the mentioned quantities in a fully consistent manner.

A separate, important field of study is the magnetocaloric effect \cite{tishin_magnetocaloric_2007,pecharsky_thermodynamics_2001,pecharsky_magnetocaloric_1999} in systems with magnetoelastic coupling  \cite{piazzi_magnetocaloric_2015,ribeiro_theoretical_2015,singh_magnetocaloric_2013,amaral_mean-field_2011,mukherjee_overcoming_2011,basso_magnetocaloric_2011,alho_influence_2010,de_oliveira_theoretical_2010,valiev_entropy_2009,plaza_magnetocaloric_2009,von_ranke_theoretical_2006,amaral_magnetoelastic_2004,reis_charge-ordering_2004}. The description of the caloric effects for the case of many interacting subsystems constitutes an interesting problem in thermodynamics and is crucial for the correct modeling of real materials. What additionally boosts the interest in magnetoelastic systems is that the presence of the coupling paves the way to tuning the magnetocaloric properties with external pressure \cite{li_hydrostatic_2017,nayak_pressure_2009,gama_pressure-induced_2004,nikitin_magnetocaloric_1991}. Therefore, microscopic models of systems with interacting lattice and magnetic components are highly desired also from magnetocaloric point of view.

In our work Ref.~\cite{balcerzak_self-consistent_2017} we have established a formalism for the description of the solid including both the lattice and the magnetic subsystem. The formalism is based on total Gibbs energy, thus it allows the determination of all the thermodynamic quantities of interest. Moreover, all the thermodynamic relations based on the appropriate derivatives of the Gibbs free energy remain fulfilled. In our previous work we have illustrated the method on the example of a solid based on s.c. lattice with the long-range interatomic interactions parametrized by Morse potential. The lattice vibrations were included in Debye form with anharmonicity introduced by variable Gr\"uneisen parameter. Moreover, the magnetic subsystem consisted of spins $S=1/2$ with nearest-neighbour ferromagnetic interactions depending on the distance between the spins according to the power law $J\propto r^{-n}$, thus providing the magnetoelastic coupling. The thermodynamics of the magnetic part was characterized within the Mean Field Approximation. The analysis of the described model shown important influence of the magnetic subsystem on the lattice properties and, in turn, the effect of the lattice subsystem on the purely magnetic behaviour. However, we have not included the external magnetic field in our previous study, limiting our interest to the phenomena related to the spontaneous magnetization. The presence of the magnetic field constitutes the next intensive thermodynamic parameter (analogous to the presence of the pressure). Therefore, in order to make the picture complete, in the present paper we investigate the thermodynamics of our model paying the attention to the magnetic field.

\section{Theoretical model}
\label{sec2}

Our system of interest is a solid based on simple cubic (s.c.) crystalline lattice with spins $S=1/2$ located at each site and interacting with nearest-neighbour ferromagnetic interaction. The thermodynamic state is described with five parameters: temperature $T$, volume $V$, pressure $p$, magnetization $m$ and external magnetic field $H$. The system volume is related to the (isotropic) relative deformation $\varepsilon$ by the formula $V=V_0\left(1+\varepsilon\right)$, where $V_0$ is the volume taken for $p=0$, $H=0$ and at $T=0$. Moreover, the convenient parameter introducing the external magnetic field is $h=-g\mu_{\rm B}H$, where $g$ is gyromagnetic factor for localized spins and $\mu_{\rm B}$ is Bohr magneton. The parameter $h$ has the dimension of energy. 

The detailed description of the model can be found in our previous work Ref.~\cite{balcerzak_self-consistent_2017}. Here, we present only a brief recapitulation for the clarity. 

The description of static lattice energy is based on long-range Morse interatomic potential:
\begin{equation}
\label{eq:Morse}
U\left(r\right)=D\left(1-e^{-\alpha\left(r-r_0\right)/r_0}\right).
\end{equation}

The lattice vibrations are included by means of anharmonic Debye model with variable Gr\"uneisen ratio, as well as with a characteristic Debye temperature $T_{\rm D}=T^{0}_{\rm D}\exp\left[\frac{3\alpha-2}{6}\frac{1-\left(1+\varepsilon\right)^{q}}{q}\right]$, where $T^{0}_{\rm D}$ is a reference Debye temperature for $\varepsilon=0$ and $h=0$ at $T=0$ and $q$ parametrizes the Debye temperature changes with volume deformations due to anharmonicity, while $\alpha$ is a Morse potential parameter. 

The description of the magnetic subsystem is performed within the Mean Field Approximation. The key assumption is the existence of distance-dependent magnetic interaction, in a form of the nearest-neighbour ferromagnetic coupling $J_1$, depending on the lattice deformation like:

\begin{equation}
\label{eq:J}
J_1=J\left(\frac{1+\varepsilon}{1+\varepsilon_{\rm C}}\right)^{-n/3},
\end{equation}
with $n$ being a decay exponent and $\varepsilon_{\rm C}$ denoting the lattice deformation at critical temperature. 

The distance (volume) dependence of the magnetic exchange integral is the source of the magnetoelastic coupling, interrelating the lattice and magnetic subsystems and causing the mutual influence of magnetic parameters on lattice ones and vice versa.

The critical (Curie) temperature is expressed within the Mean Field Approximation as $k_{\rm B}T_{\rm C}=\frac{S(S+1)}{3}z_1J_1$, where $z_1=6$ is the number of nearest neighbours in s.c. crystalline lattice. Note that $J_1$ is temperature-dependent due to its dependence on the lattice deformation $\varepsilon$, which varies with temperature.

The thermodynamic description of the system, subdivided into two interacting parts - lattice one and magnetic one - is based on the following expression for the total Gibbs energy:
\begin{equation}
\label{eq:Gibbs}
G=G_{m}+F_{\varepsilon}+F_{\rm D}+pV.
\end{equation}
In the formula above, $F_{\varepsilon}$ is the static Helmholtz free energy for the lattice subsystem (based on the Morse potential), while $F_{\rm D}$ is the Helmholtz free energy of the lattice vibrations in the anharmonic Debye approximation. The Gibbs free energy of the magnetic subsystem is denoted by $G_{m}$. The expressions for all these quantities within our approach can be found in the previous work. Ref.~\cite{balcerzak_self-consistent_2017}.

The total Gibbs energy is a function of five thermodynamic parameters: $G=G\left(\varepsilon,m,T,p,h\right)$. The variational minimization of $G$ with respect to volume deformation $\varepsilon$, where $V=V_0\left(1+\varepsilon\right)$, and magnetization $m$ leads to a pair of equations of state:
\begin{equation}
\label{eq:state1}
\frac{\partial G}{\partial V}=0
\end{equation}
and
\begin{equation}
\label{eq:state2}
\frac{\partial G}{\partial m}=0.
\end{equation}

The first equation, after taking into account Eq.~\ref{eq:state2} takes the form of:
\begin{equation}
\label{eq:state2b}
p=p_{\varepsilon}+p_{\rm D}+p_{m},
\end{equation}
where $p_{\varepsilon}=-\left(\frac{\partial F_{\varepsilon} }{\partial V}\right)_{T,h,p}$ is the contribution to the pressure originating from the static lattice deformation, $p_{\rm D}=-\left(\frac{\partial F_{\rm D} }{\partial V}\right)_{T,h,p}$ is the pressure of lattice vibrations in anharmonic Debye approximation and $p_{m}=-\left(\frac{\partial G_{m} }{\partial V}\right)_{T,h,p}$ is the magnetic contribution to the pressure. All the explicit expressions for the components of the pressure have been derived in our work Ref.~\cite{balcerzak_self-consistent_2017}. However, the entropy and the response functions have not been considered there.

The total entropy of the system can be defined as  

\begin{equation}
\label{eq:S}
S=-\left(\frac{\partial G}{\partial T}\right)_{p,h}.
\end{equation}
Taking into consideration the individual terms in the expression for the Gibbs energy ~(Eq.~\ref{eq:Gibbs}), the appropriate derivatives of lattice-related quantities can be expressed as:
\begin{equation}
\label{eq:derivative1}
\left(\frac{\partial F_{\rm \varepsilon}}{\partial T}\right)_{p,h}=-p_{\rm \varepsilon}\left(\frac{\partial V}{\partial T}\right)_{p,h}
\end{equation}
and
\begin{equation}
\label{eq:derivative2}
\left(\frac{\partial F_{\rm D}}{\partial T}\right)_{p,h}=-p_{\rm D}\left(\frac{\partial V}{\partial T}\right)_{p,h}+3Nk_{\rm B}\left[\frac{4}{y_{\rm D}^3}\frac{\pi^4}{15}+3\ln\left(1-e^{-y_{\rm D}}\right)-\frac{12}{y_{\rm D}}\mathrm{Li}_{2}\left(e^{-y_{\rm D}}\right)-\frac{24}{y_{\rm D}^2}\mathrm{Li}_{3}\left(e^{-y_{\rm D}}\right)-\frac{24}{y_{\rm D}^3}\mathrm{Li}_{4}\left(e^{-y_{\rm D}}\right)     \right].
\end{equation}

In the formula above, $y_{\rm D}=\frac{T_{\rm D}}{T}$, whereas $\mathrm Li_{s}(x)$ is the polylogarithm of order $s$ \cite{balcerzak_self-consistent_2017}. 

The temperature derivative of the magnetic Gibbs energy (\cite{balcerzak_self-consistent_2017,szalowski_phase_2008}) is:
\begin{equation}
\label{eq:Sm}
\left(\frac{\partial G_{\rm m}}{\partial T}\right)_{p,h}=-Nk_{\rm B}\ln\left\{\frac{\sinh\left[\frac{2S+1}{2}\beta\left(m\sum_{k}^{}{z_{k}J_{k}}+h\right)\right]}{\sinh\left[\frac{1}{2}\beta\left(m\sum_{k}^{}{z_{k}J_{k}}+h\right)\right]}\right\}+\frac{N}{T}m\left(m\sum_{k}^{}{z_{k}J_{k}}+h\right)-p_{m}\left(\frac{\partial V}{\partial T}\right)_{p,h}.
\end{equation}

Taking into account the equation of state (Eq.~\ref{eq:state2b}), the total entropy of the system can be finally expressed as a sum of two terms:
\begin{eqnarray}
\label{eq:Stotal}
S&=&3Nk_{\rm B}\left[\frac{4}{y_{\rm D}^3}\frac{\pi^4}{15}+3\ln\left(1-e^{-y_{\rm D}}\right)-\frac{12}{y_{\rm D}}\mathrm{Li}_{2}\left(e^{-y_{\rm D}}\right)-\frac{24}{y_{\rm D}^2}\mathrm{Li}_{3}\left(e^{-y_{\rm D}}\right)-\frac{24}{y_{\rm D}^3}\mathrm{Li}_{4}\left(e^{-y_{\rm D}}\right)     \right]\nonumber\\
&&+Nk_{\rm B}\ln\left\{\frac{\sinh\left[\frac{2S+1}{2}\beta\left(m\sum_{k}^{}{z_{k}J_{k}}+h\right)\right]}{\sinh\left[\frac{1}{2}\beta\left(m\sum_{k}^{}{z_{k}J_{k}}+h\right)\right]}\right\}-\frac{N}{T}m\left(m\sum_{k}^{}{z_{k}J_{k}}+h\right).
\end{eqnarray}

The first term in the above formula can be identified with the lattice subsystem, while the remaining part is related to the magnetic part. It should be strongly emphasized that, in spite of this subdivision, all the terms depend on all thermodynamic parameters. This is due to the fact that all thermodynamic parameters - $T,p,V,h,m$ - are interrelated by a pair of equations of state (Eqs.~\ref{eq:state1} and {\ref{eq:state2}). Therefore, for example the pressure has an impact on the magnetization, while the magnetic field influences the volume.

On the basis of the entropy and its components assigned to particular subsystem, further thermodynamic quantities can be defined. The important and directly measurable quantities are specific heats. For systems of purely lattice character the usual choices are: specific heat for constant pressure $p$ or for constant volume $V$. However, when the system is described with the parameters $p,V,h,m$ at given temperature, specific heats can be determined at two constant thermodynamic parameters. A natural selection is the fixed value of external magnetic field $h$. Therefore, two specific heats can be of special interest:
\begin{equation}
\label{eq:cv}
c_{V,h}=T\left(\frac{\partial S}{\partial T}\right)_{V,h}
\end{equation}
and 
\begin{equation}
\label{eq:cp}
c_{p,h}=T\left(\frac{\partial S}{\partial T}\right)_{p,h}.
\end{equation}
In analogy to the entropy (Eq.~\ref{eq:Stotal}), each of the specific heats can be additionally separated into a contribution of terms attributed to the lattice and the magnetic subsystem.

The knowledge of the entropy as a function of temperature, pressure and external magnetic field gives an opportunity to study the magnetocaloric effect in the system \cite{tishin_magnetocaloric_2007,pecharsky_thermodynamics_2001,pecharsky_magnetocaloric_1999,szalowski_thermodynamic_2011}. Such effect consists in the change of system entropy under isothermal change of external magnetic field (what implies exchange of heat between the system and the environment) or system temperature change under adiabatic variation of the magnetic field. The first aspect can be quantitatively characterized by introducing isothermal entropy change $\Delta S_{T,p}=S\left(T,p,h=0\right)-S\left(T,p,h\right)$ for the field variation between $h=0$ and $h>0$. The second aspect can be described by means of an adiabatic temperature change for field varying between $h=0$ and $h>0$, defined by the equation $S\left(T,p,h\right)=S\left(T-\Delta T_{S,p},p,h=0\right)$.

The presence of the magnetoelastic coupling in the system permits the analysis of the influence of external magnetic field on the lattice response functions. The most common response functions related to the lattice properties are: the thermal expansion coefficient 
\begin{equation}
\label{eq;alpha}
\alpha_{p,h}=\frac{1}{V}\left(\frac{\partial V}{\partial T}\right)_{p,h}
\end{equation}
and the isothermal compressibility
\begin{equation}
\label{eq:kappa}
\kappa_{T,h}=-\frac{1}{V}\left(\frac{\partial V}{\partial p}\right)_{T,h}.
\end{equation}
In addition, the most pronounced expected consequence of the magnetoelastic coupling is the lattice deformation under the influence of the magnetic field. In order to quantify this effect, a magnetostriction coefficient can be defined as 
\begin{equation}
\label{eq:lambda}
\lambda_{T,p}=\frac{1}{V}\left(\frac{\partial V}{\partial h}\right)_{T,p}.
\end{equation}

It is also quite interesting to relate the magnetostriction coefficient with another quantity describing the consequences of the magnetoelastic coupling. Namely, exploiting the relation $\frac{\partial^2 G}{\partial p\partial h}=\frac{\partial^2 G}{\partial h\partial p}$, it can be shown that $\left(\frac{\partial V}{\partial h}\right)_{T,p}=-\left(\frac{\partial m}{\partial p}\right)_{T,h}$. The right hand side of the equality describes the piezomagnetic effect, i.e. the sensitivity of the magnetization to the external pressure. As it is seen, this quantity is directly related to the magnetostriction coefficient.

\section{Numerical results and discussion}
\label{sec3}

In order to illustrate the consequences of the magnetoelastic coupling, we have performed the numerical calculations of the above mentioned thermodynamic quantities, putting emphasis mainly on the influence of the magnetic field and of pressure.

\subsection{Lattice and magnetic response functions}

\begin{figure}[h!]
  \begin{center}
\includegraphics[scale=0.40]{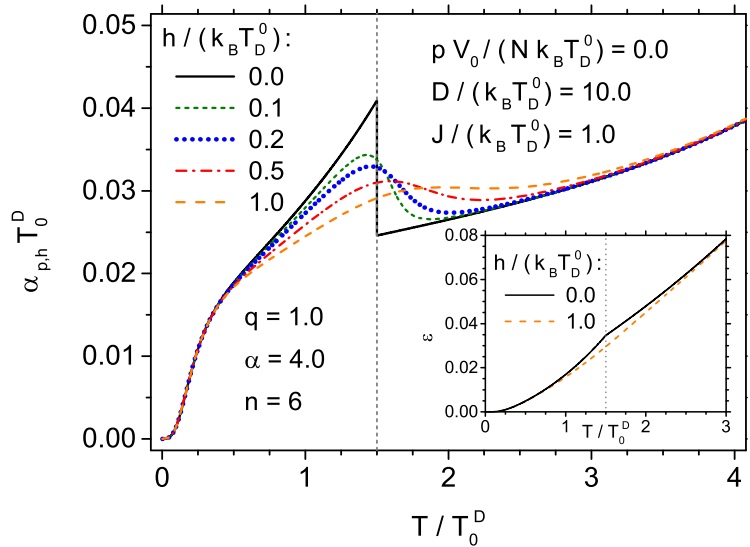}
  \end{center}
   \caption{\label{fig:alpha} Reduced thermal expansion coefficient at constant pressure and magnetic field, as a function of reduced temperature for various values of the external magnetic field. The dashed vertical line indicates the critical temperature in the absence of the field. The inset shows the isotropic volume deformation as a function of the reduced temperature in the absence and in the presence of strong external magnetic field.}
\end{figure}

\begin{figure}[h!]
  \begin{center}
\includegraphics[scale=0.40]{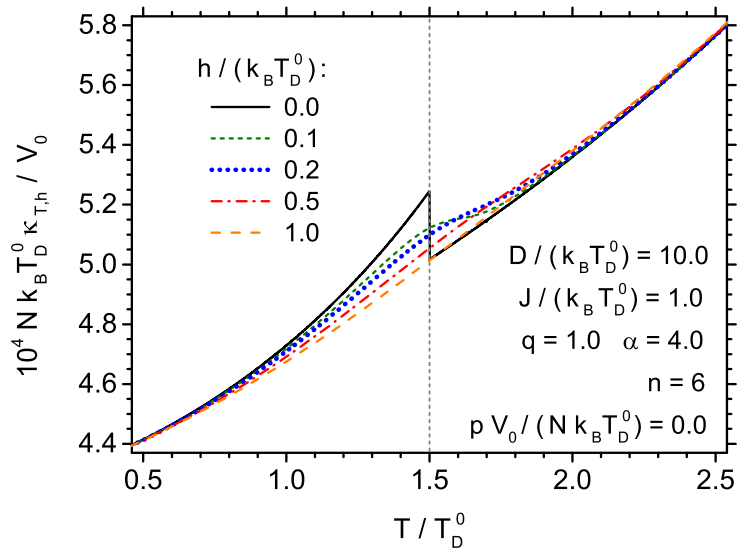}
  \end{center}
   \caption{\label{fig:kappa} Reduced isothermal compressibility at constant temperature and magnetic field as a function of the reduced temperature. Various values of external magnetic field are assumed. The dashed vertical line indicates the critical temperature in the absence of the field.  }
\end{figure}

\begin{figure}[h!]
  \begin{center}
\includegraphics[scale=0.40]{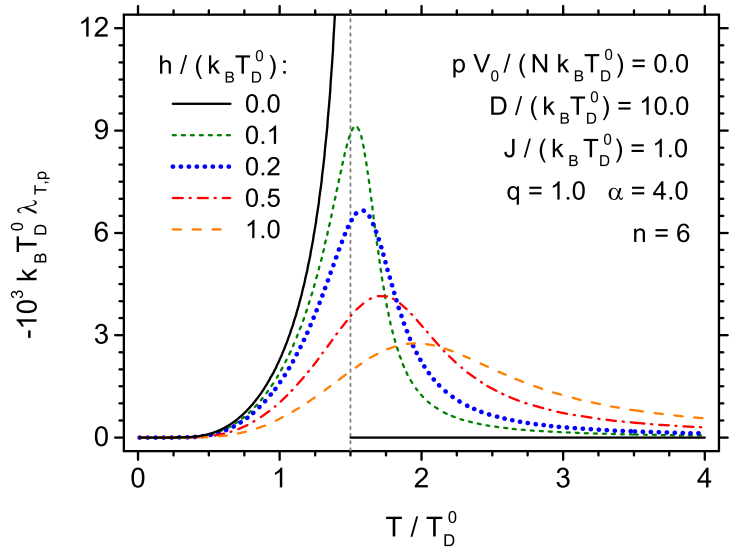}
  \end{center}
   \caption{\label{fig:lambda} Reduced magnetostriction coefficient at constant temperature and pressure, as a function of reduced temperature, for various values of external magnetic field. The dashed vertical line indicates the critical temperature in the absence of the field. }
\end{figure}

\begin{figure}[h!]
  \begin{center}
\includegraphics[scale=0.40]{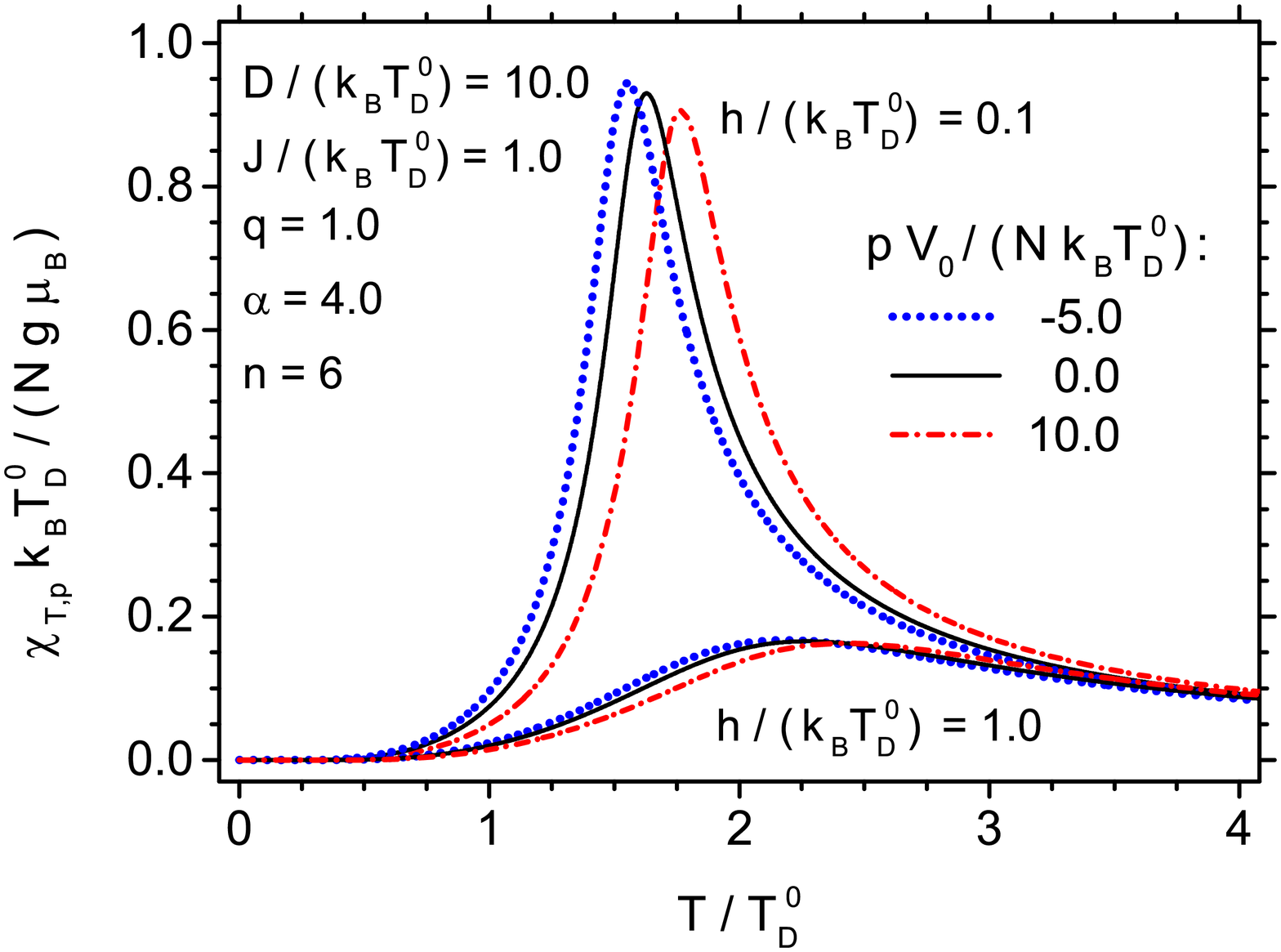}
  \end{center}
   \caption{\label{fig:chi} Reduced magnetic susceptibility at constant temperature and pressure, as a function of reduced temperature. The cases of zero pressure, a positive and a negative pressure are illustrated for weak and strong external magnetic field. }
\end{figure}

The coupling between lattice and magnetic subsystems manifests itself in the influence of the external magnetic field on purely lattice properties. Therefore, it is worthy of analysis how the fundamental response functions of the system related to the non-magnetic properties are modified under the action of the magnetic field.

One of the examples can be volume expansion coefficient calculated for constant pressure and magnetic field. The temperature dependence of such response function, describing the lattice system response to temperature variations,  can be followed in Fig.~\ref{fig:alpha}, which emphasizes the effect of the external magnetic field (up to $h/\left(k_{\rm B}T_{\rm D}^{0}\right)$=1.0). As it was discussed in our work Ref.~\cite{balcerzak_self-consistent_2017}, the expansion coefficient exhibits a discontinuous behaviour at the critical temperature due to magnetic phase transition. However, in the presence of the external magnetic field no phase transition takes place in our system (as the magnetization does not vanish completely at any temperature). Therefore, a continuous behaviour of $\alpha_{p,h}$ can be expected for $h>0$. Such a qualitative change of expansion coefficient thermal dependence can be seen in Fig.~\ref{fig:alpha}. For $h>0$ the jump is replaced with a maximum which, for increasing $h$, flattens and becomes displaced towards higher temperatures. Therefore, for strong magnetic field, the thermal dependence of expansion coefficient resembles the one predicted for pure lattice system, without magnetic properties. The decrease of $\alpha_{p,h}$ in the external magnetic field is also visible directly in the thermal behaviour of relative volume deformation $\varepsilon$, as illustrated in the inset in Fig.~\ref{fig:alpha}, contrasting the case of $h/\left(k_{\rm B}T_{\rm D}^{0}\right)$=0.0 and 1.0. Clearly, the relative deformation at given temperature is decreased by the external field. 

Another lattice-related response function, indicating the discontinuous behaviour at the magnetic phase transition, is the isothermal compressibility. Fig.~\ref{fig:kappa} allows the discussion of the influence of magnetic field on that quantity. In similar manner as in the case of $\alpha_{p,h}$, the presence of the magnetic field causes the discontinuity to vanish and the value of $\kappa_{T,h}$ is reduced for $T<T_{\rm C}$ while it increases for $T>T_{\rm C}$. The jump is first transformed into a weakly pronounced maximum, which flattens and, for even stronger fields, the thermal dependence becomes strictly monotonous. Once more, the strong magnetic field restores the behaviour characteristic of pure lattice system. 

Due to the presence of the magnetoelastic coupling, another response function can be introduced, which is specific for such composite system. Namely, the influence of the variation of magnetic field on the system volume can be characterized with the appropriate response function, being a magnetostriction coefficient. Let us illustrate this quantity in Fig.~\ref{fig:lambda}, as dependent on the temperature, for various external magnetic fields. Let us note here that this coefficient has negative sign, as the volume deformation decreases under the action of $h$ (see inset in Fig.~\ref{fig:alpha}). At the field $h=0$, the sensitivity of the system volume to the differential increase of the field rises when the temperature varies between 0 and $T_{\rm C}$, showing a discontinuity at the critical temperature. On the contrary, for $T>T_{\rm C}$ this quantity is equal to 0. In the finite magnetic field $\lambda_{p,T}$ shows a continuous behaviour with a maximum close to $T_{\rm C}$. With the increase of $h$ the magnitude of magnetostriction coefficient is reduced fast and the position of the maximum is shifted towards higher temperatures. The discussed quantity is only characteristic of systems with magnetoelastic coupling and has no correspondence in purely lattice system. It should be mentioned that the product $-V\lambda_{p,T}$ is equal to the derivative $\left(\partial m/\partial p\right)_{T,h}$, which quantifies the piezomagnetic effect.

It is evident that the external magnetic field influences the lattice-related response functions in a system with magnetoelastic coupling. In turn, the pressure should modify the response functions related to the magnetic subsystem. Such a response function is the magnetic susceptibility. Its temperature dependence is plotted in Fig.~\ref{fig:chi} for $p=0$ as well as for selected positive and negative pressures. Moreover, the presence of a weak or a strong magnetic field is assumed. In a weak field, $\chi_{T}$ exhibits a pronounced maximum at the temperature slightly larger than the critical temperature (defined for $h=0$). This maximum is sensitive to the pressure, as the positive pressure reduces the magnitude of susceptibility at the maximum. The position of the maximum follows the pressure dependence of the critical temperature (which increases with the increase in $p$). The negative pressure exerts an opposite effect. When the strong magnetic field is present, the maximum of susceptibility has greatly reduced magnitude, significant width and is more shifted towards higher temperatures with respect to $T_{\rm C}$. However, also in this case the effect of the pressure can be noticed in analogy to the case of the weak field (but is less visible due to the overall reduced susceptibility values). It should be noted that Mean Field Approximation predicts the infinite susceptibility at $T=T_{\rm C}$ in the absence of the magnetic field (not shown in Fig.~\ref{fig:chi}) and also in this case the position of the discontinuity follows the pressure dependence of $T_{\rm C}$. 

\subsection{Entropy and specific heat}

\begin{figure}[h!]
  \begin{center}
\includegraphics[scale=0.40]{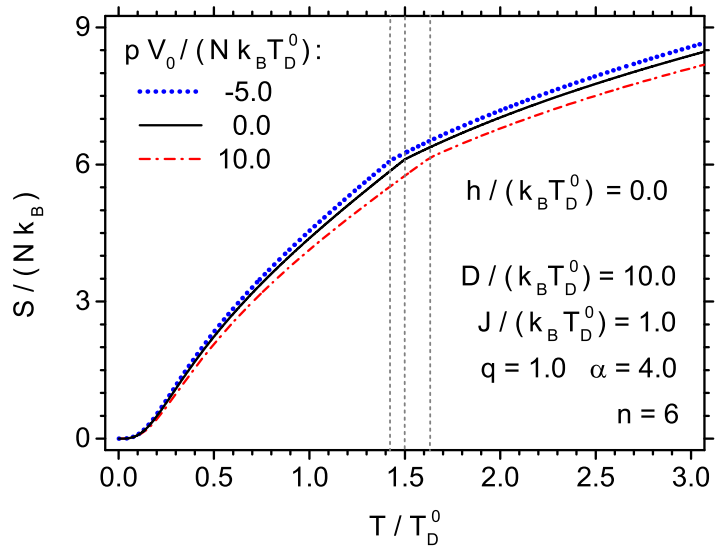}
  \end{center}
   \caption{\label{fig:Sp} The total entropy of the system as a function of reduced temperature. The calculations are performed for zero pressure as well as for a negative and positive pressure. The dashed vertical lines indicate the critical temperatures in the absence of external magnetic field for each considered pressure value.}
\end{figure}

\begin{figure}[h!]
  \begin{center}
\includegraphics[scale=0.40]{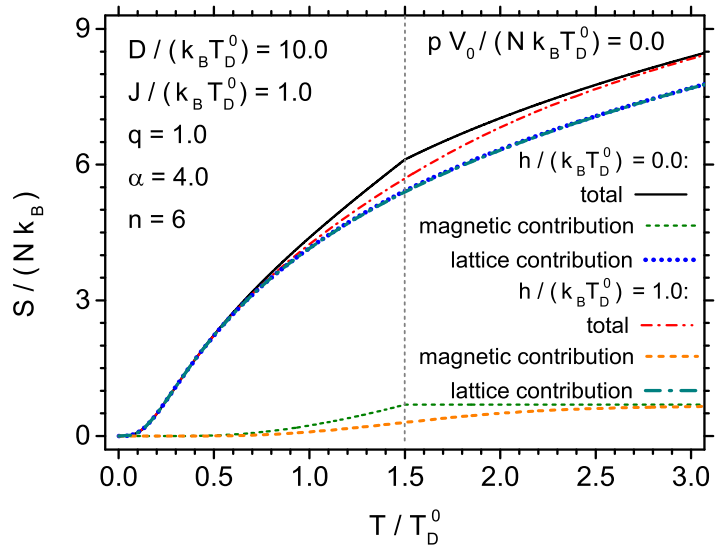}
  \end{center}
   \caption{\label{fig:Sh} The total entropy as well as the contributions of the terms in Eq.~\ref{eq:Stotal} attributed to lattice and magnetic subsystems as a function of reduced temperature. The cases of zero magnetic field and strong magnetic field are shown. The dashed vertical line indicates the critical temperature in the absence of the field. }
\end{figure}

\begin{figure}[h!]
  \begin{center}
\includegraphics[scale=0.40]{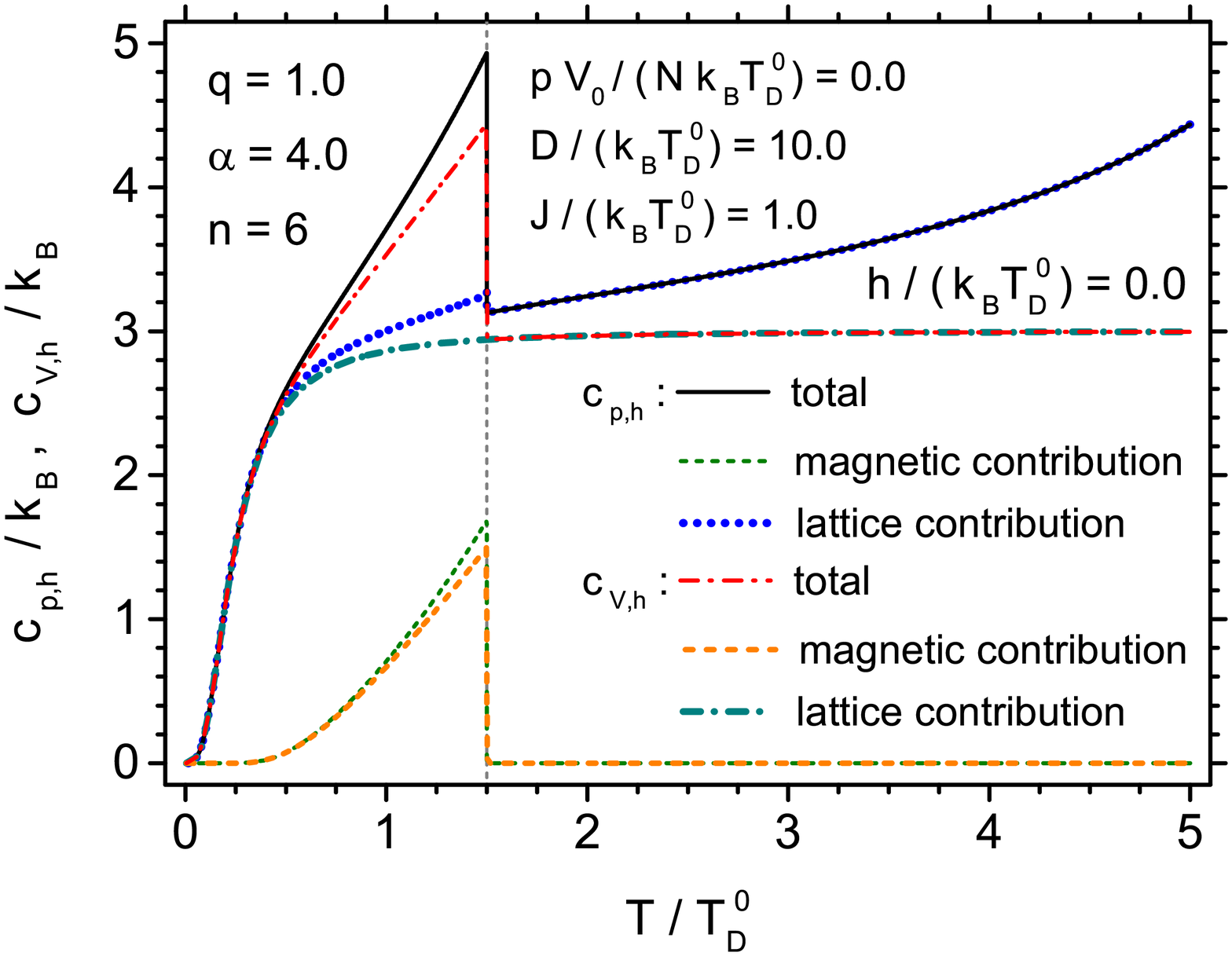}
  \end{center}
   \caption{\label{fig:CpV1} Specific heats at constant pressure and volume and constant magnetic field as a function of reduced temperature, calculated in the absence of the magnetic field. The total specific heat is decomposed into contributions of two terms attributed to the lattice and the magnetic subsystems. The dashed vertical line indicates the critical temperature in the absence of the field.}
\end{figure}

\begin{figure}[h!]
  \begin{center}
\includegraphics[scale=0.40]{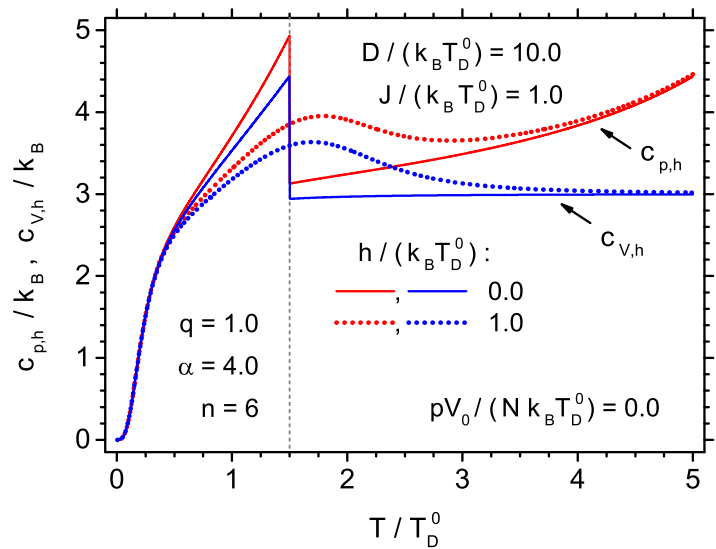}
  \end{center}
   \caption{\label{fig:cpVH} Specific heats at constant pressure and volume and constant magnetic field as a function of reduced temperature, compared in the absence and in the presence of strong magnetic field. The dashed vertical line indicates the critical temperature in the absence of the field.}
\end{figure}

\begin{figure}[h!]
  \begin{center}
\includegraphics[scale=0.40]{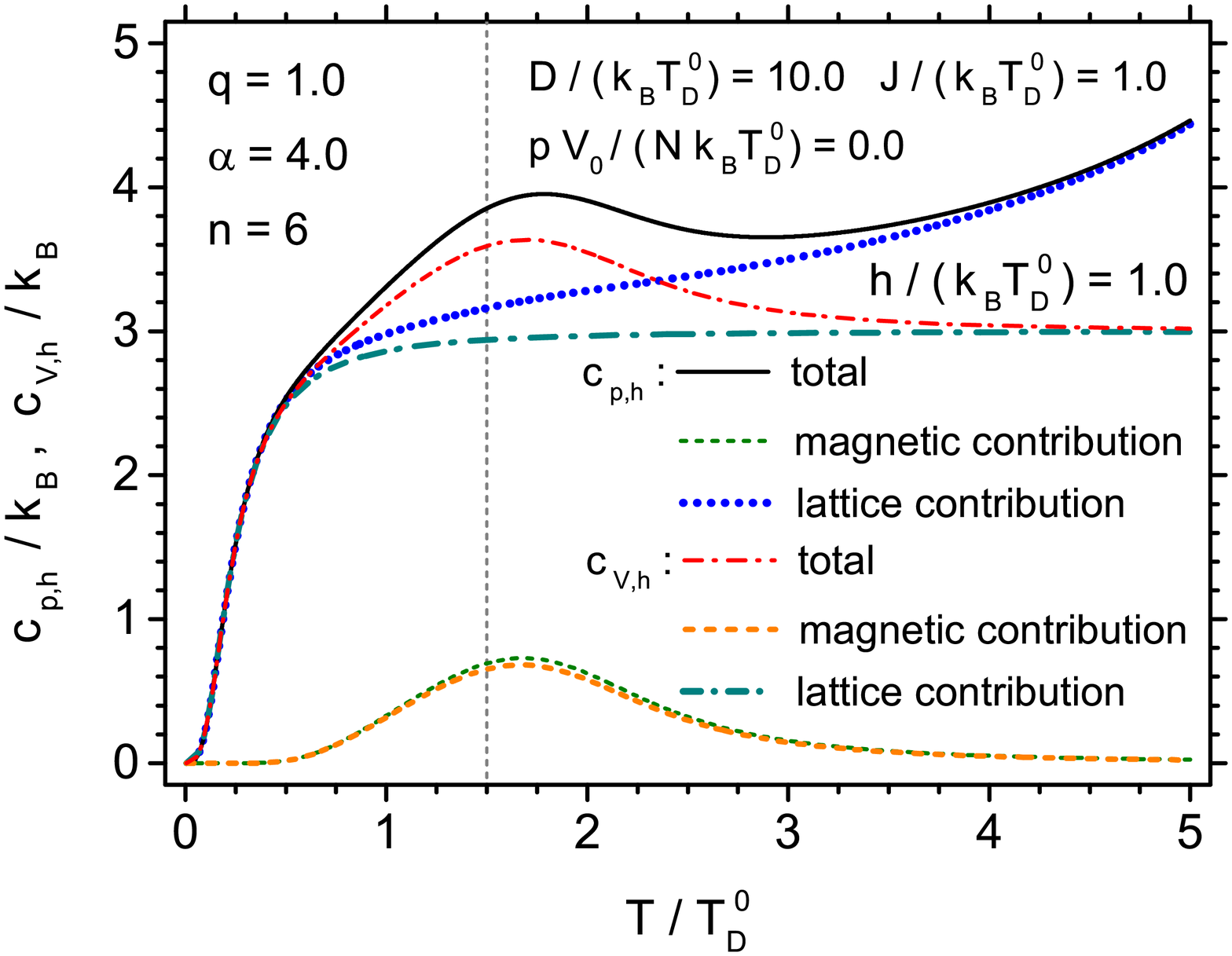}
  \end{center}
   \caption{\label{fig:CpV2} Specific heats at constant pressure and volume and constant magnetic field as a function of reduced temperature, calculated in the presence of a strong  magnetic field. The total specific heat is decomposed into contributions of two terms attributed to the lattice and the magnetic subsystems. The dashed vertical line indicates the critical temperature in the absence of the field.}
\end{figure}

\begin{figure}[h!]
  \begin{center}
\includegraphics[scale=0.40]{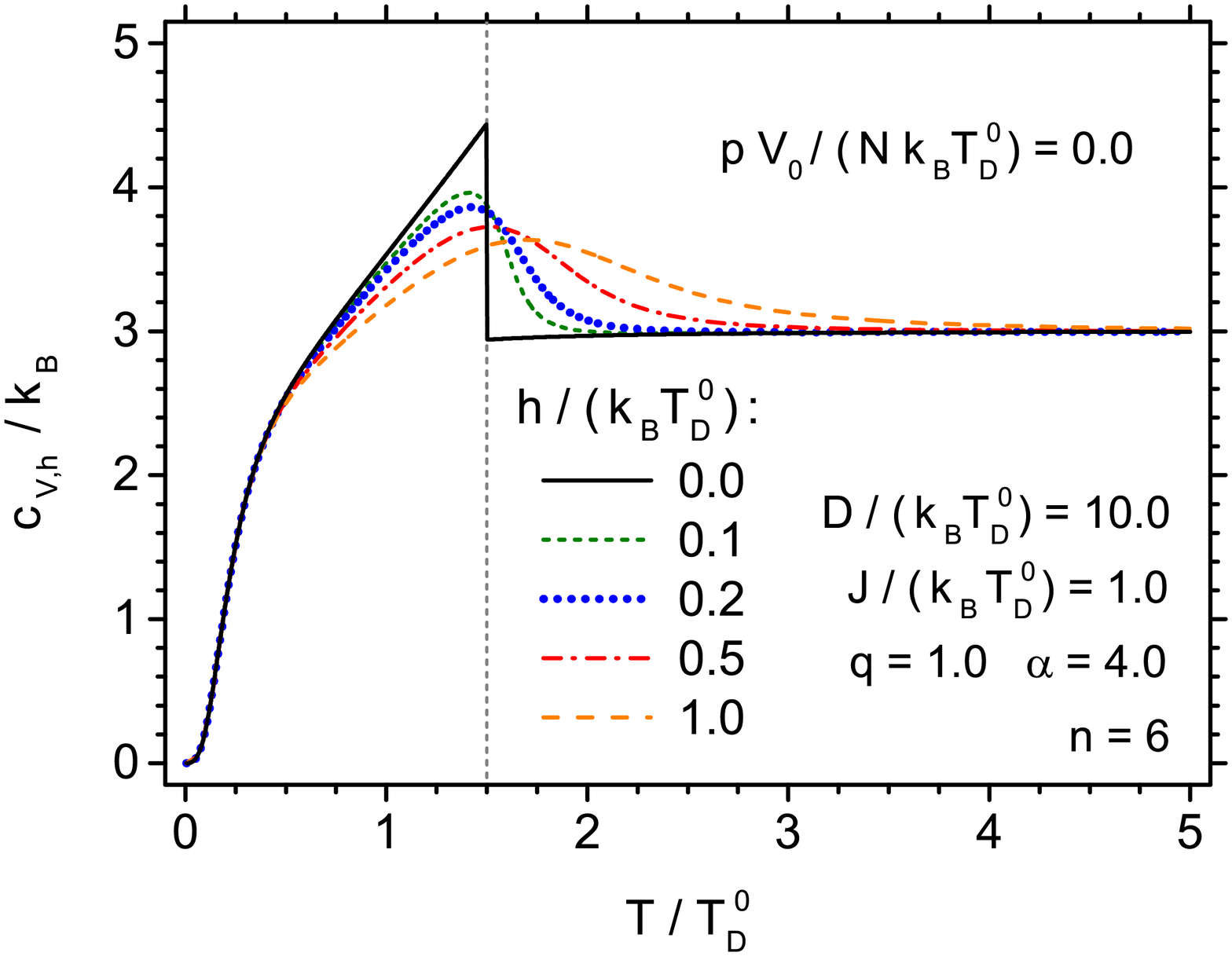}
  \end{center}
   \caption{\label{fig:cV} Specific heat at constant volume and magnetic field as a function of the reduced temperature, plotted for various values of external magnetic field. The dashed vertical line indicates the critical temperature in the absence of the field.}
\end{figure}

\begin{figure}[h!]
  \begin{center}
\includegraphics[scale=0.40]{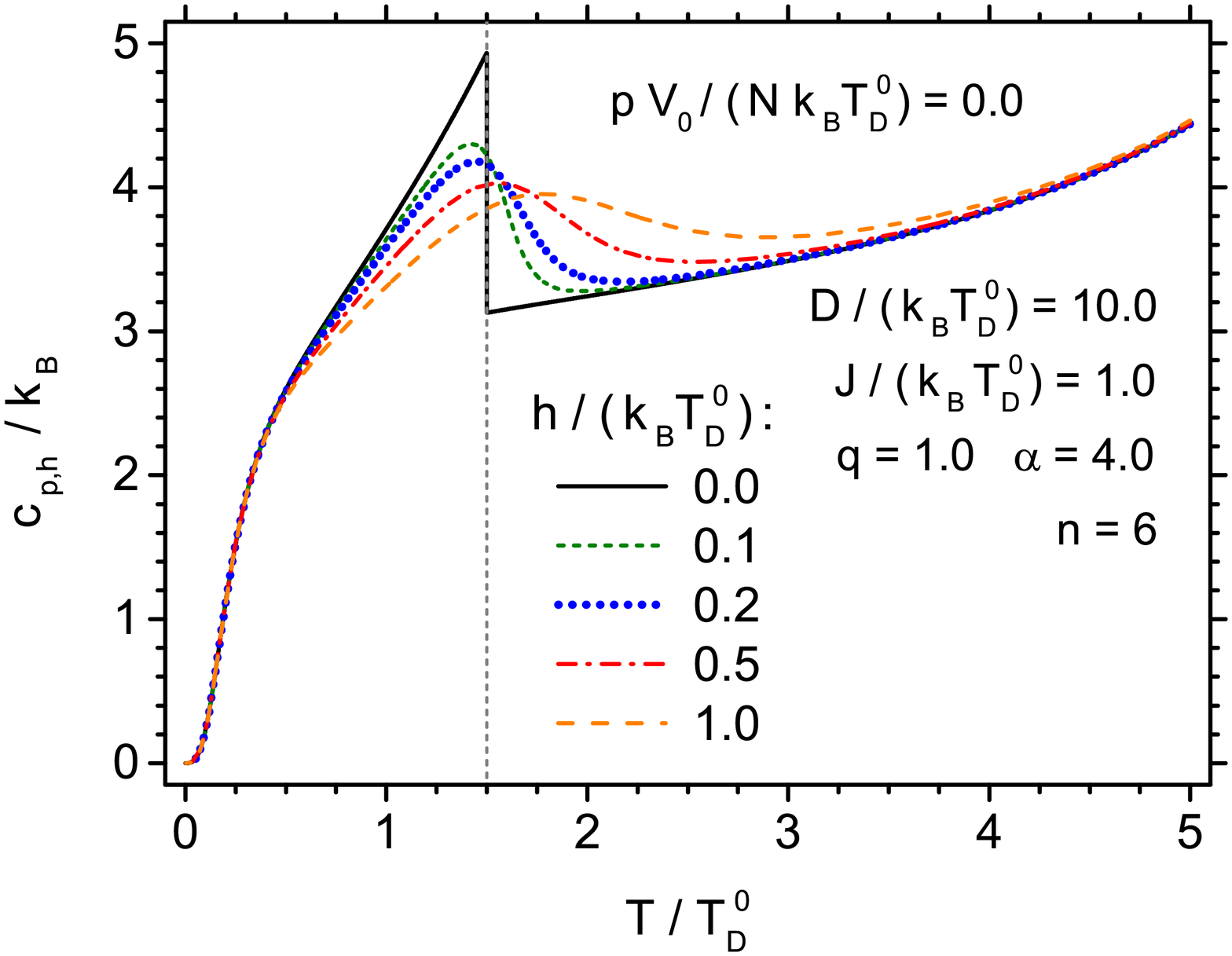}
  \end{center}
   \caption{\label{fig:cp} Specific heat at constant pressure and magnetic field as a function of the reduced temperature, plotted for various values of external magnetic field. The dashed vertical line indicates the critical temperature in the absence of the field.}
\end{figure}

The important quantity characterizing the thermodynamics of the system is its entropy, which in the present case is sensitive both to the pressure and magnetic field. The description of derivatives of the entropy enables the characterization of the system response to the temperature change (calculation of the specific heats). Moreover, the knowledge of entropy is vital for discussion of the magnetocaloric effect. The temperature dependence of entropy can be followed in Fig.~\ref{fig:Sp}, where the total entropy of the system in question is plotted for $p=0$ as well as for selected positive and negative pressures, in the absence of the external magnetic field. The entropy is a continuous function of the temperature, exhibiting a kink at the temperature of the magnetic phase transition (the critical temperatures for three considered pressures are indicated with vertical dashed lines). The positive pressure tends to decrease the total entropy and shifts the kink towards higher temperatures (as it shifts the critical temperature of the magnetic phase transition). As the entropy can be subdivided into lattice and magnetic subsystem contributions, it is instructive to analyse the thermal dependence of the two components, both in the presence and in the absence of the magnetic field, as it is shown in Fig.~\ref{fig:Sh}. First, it can be observed that the purely lattice entropy exhibits a weak dependence on the external magnetic field. The magnetic entropy for zero field increases for $T<T_{\rm C}$ and saturates further (which is a result of the Mean Field Approximation, neglecting the magnetic correlations). The kink in magnetic entropy is reflected in the analogous behaviour of the total entropy. The influence of the external magnetic field on the magnetic entropy consists in decreasing its value, so that the magnetic entropy becomes a smoothly increasing function of the temperature in the whole range of temperatures. This reduction implies also the reduction of the total entropy. 

It should be underlined that the importance of the magnetoelastic coupling and the mutual interrelations between lattice and magnetic subsystem are much more emphasized in the thermodynamic functions being the derivatives of the entropy than in the entropy itself. Therefore, we analyse in a detailed way the behaviour of the specific heat. The usually considered specific heats for lattice systems include specific heat for constant volume or pressure. Bearing in mind that our system in question is controlled additionally with external magnetic field, we can also set this quantity as a constant. It should be mentioned that the specific heat calculated under the assumption of constant magnetization would cause the contribution from the magnetic system to vanish rigorously. Let us commence the analysis from the comparison of the specific heats $c_{p,h}$ and $c_{V,h}$ in the absence of the magnetic field, as shown in Fig.~\ref{fig:CpV1}. In the plot the total specific heat is subdivided into the contributions of lattice and magnetic subsystem. Both $c_{p,h}$ and $c_{V,h}$ indicate a discontinuity at $T=T_{\rm C}$ due to magnetic phase transition. However, let us observe, that the purely magnetic contributions differ for constant $p$ and constant $V$ (the latter one being slightly reduced in magnitude). Moreover, it should be noticed that also the lattice contribution itself shows a discontinuity for the case of $c_{p,h}$ (while it is continuous for $c_{V,h}$). Due to neglect of magnetic correlations for $T>T_{\rm C}$ by the Mean Field Approximation, the magnetic contribution vanishes for that range of temperatures. Of course, always $c_{p,h}>c_{V,h}$ (which can be also followed in Fig.~\ref{fig:cpVH} both in the absence and in the presence of the magnetic field). Due to anharmonic character of lattice vibrations and the deformation dependence of the Debye temperature the specific heat $c_{p,h}$ increases further in the high temperature range (owing to the behaviour of the lattice contribution). On the other hand, the lattice contribution for constant $V$ saturates at high temperatures at the value of $3k_{\rm B}$, in accord with the Dulong-Petit rule. The analogous analysis can be preformed in the presence of the magnetic field (Fig.~\ref{fig:CpV2}). In such case the magnetic contribution to the specific heat is continuous and shows a single maximum shifted towards temperatures higher than $T_{\rm C}$, what implies a similar maximum in the total specific heat. A slight difference between the magnetic contribution values at constant $p$ and $V$ is conserved. The behaviour of the lattice contributions is also continuous in all the cases.

Let us comment that the presence of the magnetic field at high temperatures promotes the magnetization which enters the expression for the magnetic internal energy and for the magnetic specific heat. Therefore, shifting of the specific heat maximum towards higher temperatures is observed also in purely magnetic models. A similar shifting effect is transferred to the lattice-related quantities via the magnetoelastic coupling in the present model.

The detailed response of $c_{V,h}$ and $c_{p,h}$ to the external magnetic field can be traced in Figs.~\ref{fig:cV} and \ref{fig:cp}, respectively. The transformation of the discontinuity into a maximum with increasing width and decreasing height, moving away from the critical temperature, can be seen clearly. Of course the sensitivity of the specific heat to magnetic field is highest at temperatures close to the critical temperature. Let us observe that the temperature dependence of $c_{p,h}$ bears some qualitative resemblance to the analogous dependence of thermal expansion coefficient $\alpha_{p,h}$. This is caused by the fact that the specific heat $c_{p,h}$ contains the derivative of entropy with respect to temperature (calculated at constant $p$ and $h$). As the entropy is a function of $T$ and $\varepsilon$, its derivative depends on $\varepsilon$ as well as on $(\partial \varepsilon/\partial T)_{p,h}$, while the latter quantity is proportional to $\alpha_{p,h}$. These facts explain the similarity noticed above.

\subsection{Magnetocaloric effect}

\begin{figure}[h!]
  \begin{center}
\includegraphics[scale=0.40]{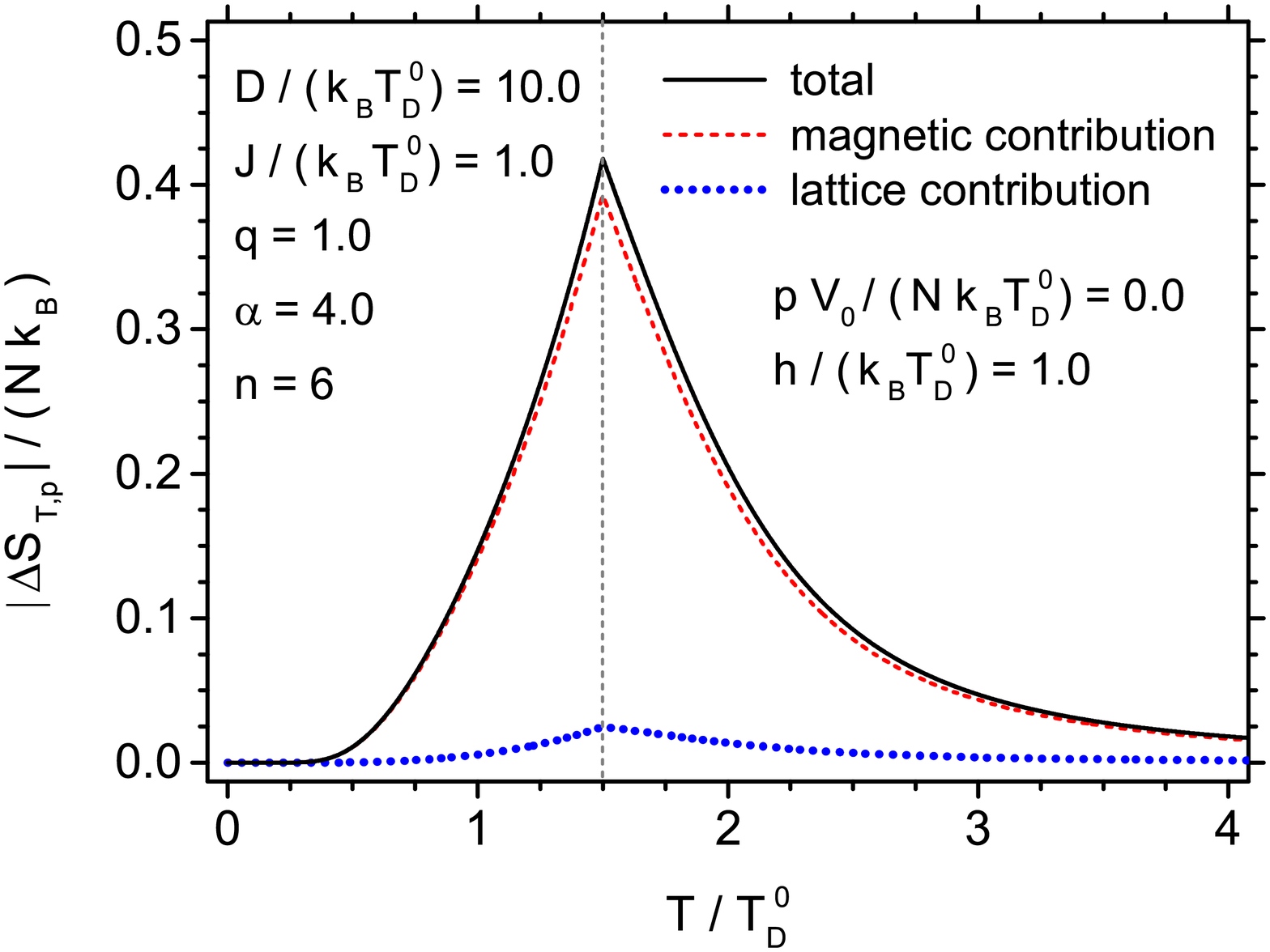}
  \end{center}
   \caption{\label{fig:deltaST1} The absolute value of isothermal entropy change at constant pressure, for the change of the external magnetic field between 0 and given value $h$, as a function of reduced temperature. The contributions originating from the two terms in Eq.~\ref{eq:Stotal} attributed to the lattice and the magnetic subsystem are separated. The dashed vertical line indicates the critical temperature in the absence of the field.}
\end{figure}

\begin{figure}[h!]
  \begin{center}
\includegraphics[scale=0.40]{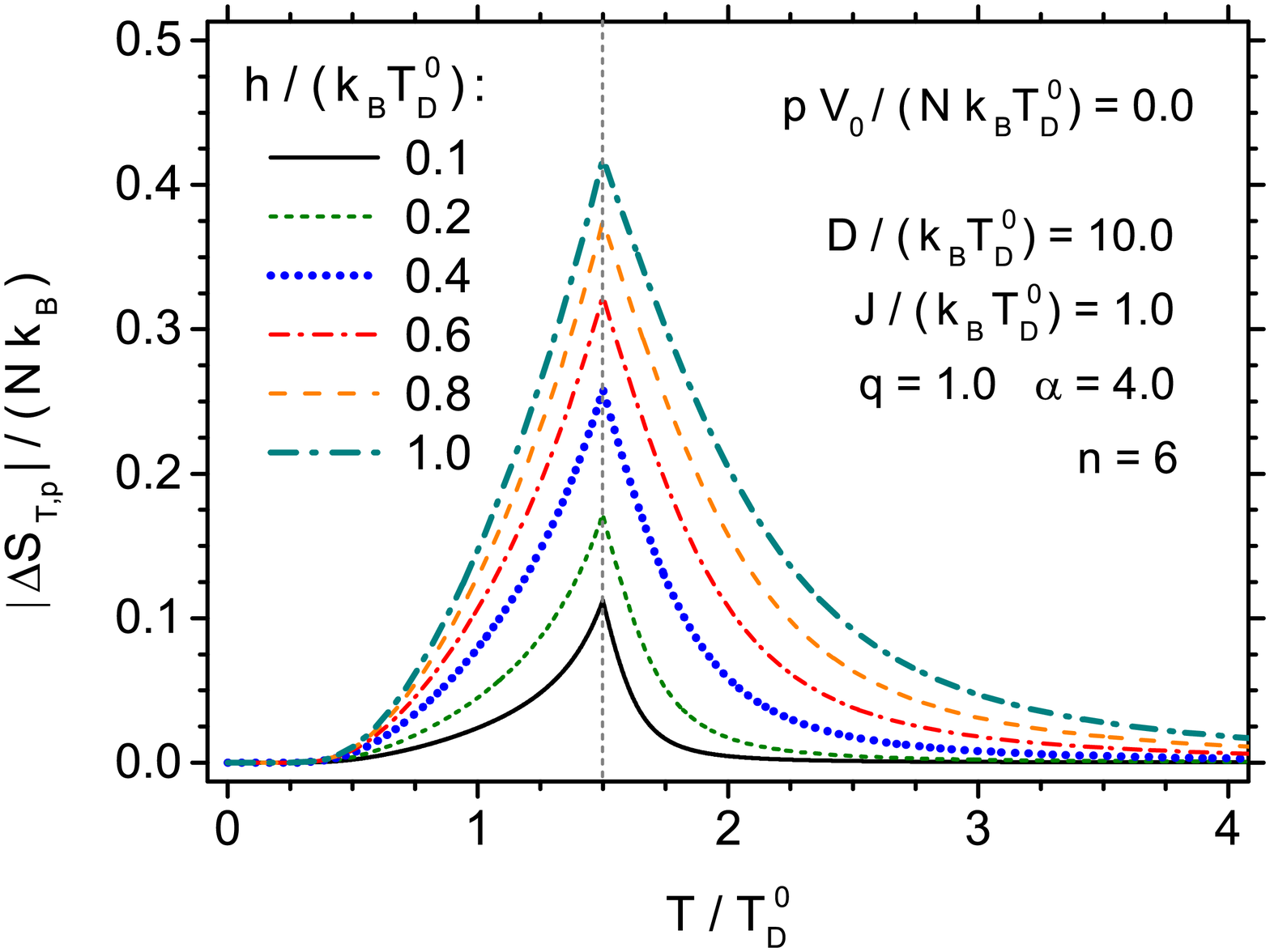}
  \end{center}
   \caption{\label{fig:deltaST2} The absolute value of isothermal entropy change at constant pressure, for the change of the external magnetic field between 0 and given value $h$, as a function of reduced temperature. The results for various amplitudes of magnetic field $h$ are shown. The dashed vertical line indicates the critical temperature in the absence of the field.}
\end{figure}

\begin{figure}[h!]
  \begin{center}
\includegraphics[scale=0.40]{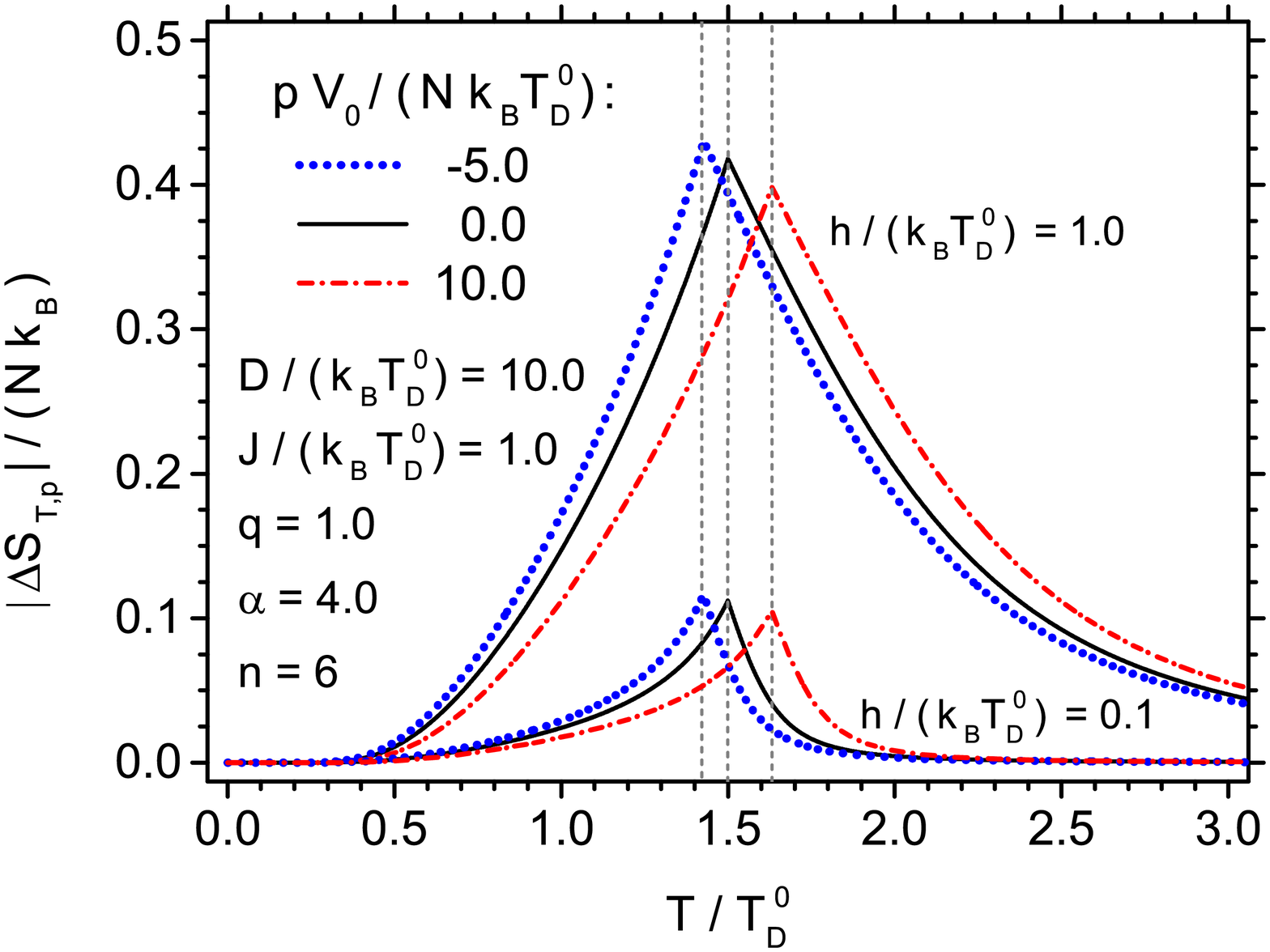}
  \end{center}
   \caption{\label{fig:deltaST3} The absolute value of isothermal entropy change at constant pressure, for the change of the external magnetic field between 0 and given value $h$, as a function of reduced temperature. The cases of zero pressure, a positive and a negative pressure are illustrated for weak and strong external magnetic field. The dashed vertical lines indicate the critical temperatures in the absence of external magnetic field for each considered pressure value. }
\end{figure}

\begin{figure}[h!]
  \begin{center}
\includegraphics[scale=0.40]{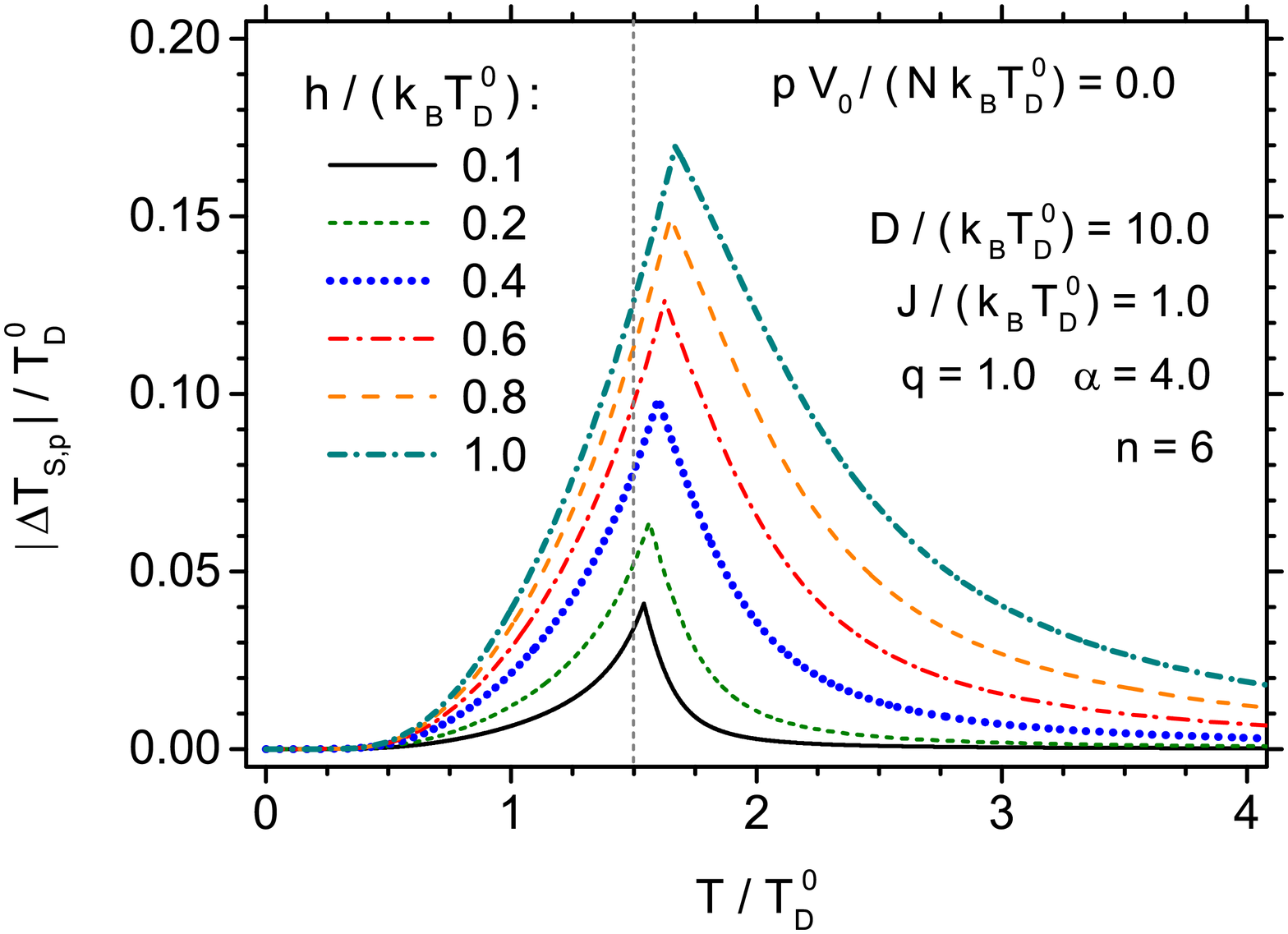}
  \end{center}
   \caption{\label{fig:deltaTS1} The absolute value of adiabatic temperature change at constant pressure, for the change of the external magnetic field between 0 and given value $h$, as a function of reduced temperature. The results for various amplitudes of magnetic field $h$ are shown. The dashed vertical line indicates the critical temperature in the absence of the field.  }
\end{figure}

\begin{figure}[h!]
  \begin{center}
\includegraphics[scale=0.40]{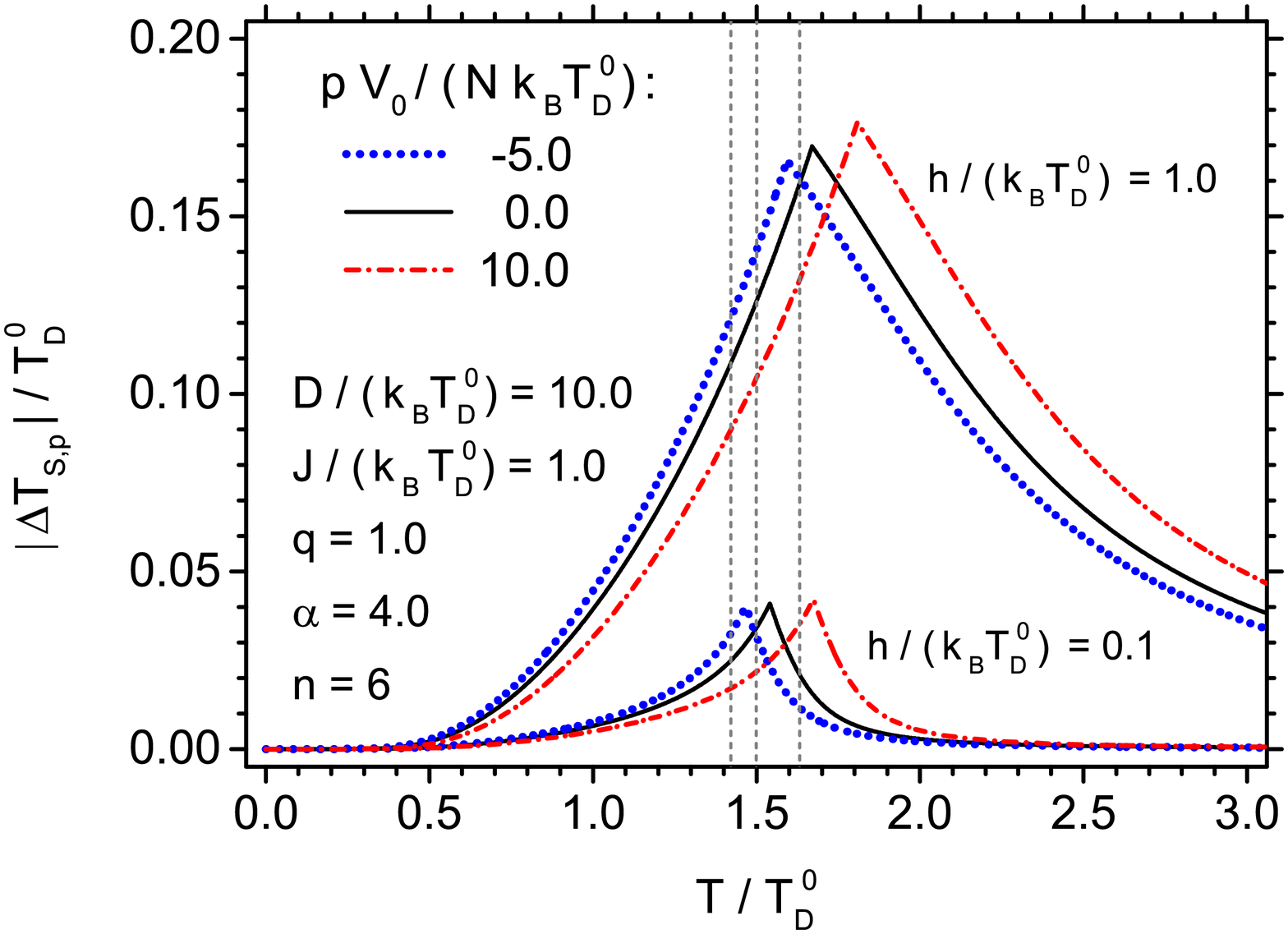}
  \end{center}
   \caption{\label{fig:deltaTS2} The absolute value of adiabatic temperature change at constant pressure, for the change of the external magnetic field between 0 and given value $h$, as a function of reduced temperature.The cases of zero pressure, a positive and a negative pressure are illustrated for weak and strong external magnetic field. The dashed vertical lines indicate the critical temperatures in the absence of external magnetic field for each considered pressure value. }
\end{figure}

One of the important phenomena related to the influence of the external magnetic field on the magnetic system is the magnetocaloric effect. Its possible manifestation is the isothermal heat exchange when the field is varying between zero and non-zero value. The quantitative measure of this aspect is the isothermal entropy change for a specified amplitude of external magnetic field variation. Another manifestation is the adiabatic change in the system temperature when the external magnetic field varies between zero and a non-zero value.

It is interesting to follow the variability of $\left|\Delta S_{T,p}\right|$ as a function of temperature at which the magnetic field changes take place, for a fixed value of amplitude of magnetic field. A representative plot is presented in Fig.~\ref{fig:deltaST1} for some large field amplitude $h/\left(k_{\rm B}T_{\rm D}^{0}\right)$=1.0. In addition to the total entropy change, also the contributions of the magnetic entropy change and the lattice entropy change are separated. It is visible that the isothermal entropy change shows a single distinct maximum, rather symmetric in shape with respect to the critical temperature (indicated with dashed vertical line). Apart form the dominant contribution of the purely magnetic entropy change, also the lattice entropy change follows the same variability pattern with a maximum at $T_{\rm C}$, however, its magnitude amounts to approximately a few per cent of the magnitude of the magnetic contribution.

An important factor shaping the isothermal entropy change is the amplitude of the magnetic field. Its influence can be followed in Fig.~\ref{fig:deltaST2} for a wide range of normalized fields $h/\left(k_{\rm B}T_{\rm D}^{0}\right)$ = 0.1 $\dots$ 1.0. In all the cases a single maximum is present and its location is exactly at the critical temperature. For lower field amplitudes the maxima exhibit asymmetric shapes, as $\left|\Delta S_{T,p}\right|$ decreases faster for $T>T_{\rm C}$. When $h$ increases, a more symmetric behaviour is observed and $\left|\Delta S_{T,p}\right|$ even tends to decrease slower for $T>T_{\rm C}$. Regarding the maximum magnitudes of isothermal entropy changes (at $T=T_{\rm C}$), a non-linear variability as a function of $h$ is clearly noticeable, as the increase is much faster for lower amplitudes of the magnetic field. 

The presence of the magnetoelastic coupling between the magnetic and lattice subsystems allows us to predict the influence of the pressure on the quantities characterizing the magnetocaloric effect. Such influence is illustrated in Fig.~\ref{fig:deltaST3}. Let us recall that the critical (Curie) temperature within our model is also pressure-dependent and is an increasing function of $p$ (see for example Fig.~1 in our work Ref.~\cite{balcerzak_self-consistent_2017}). The temperature dependences of $\left|\Delta S_{T,p}\right|$ for the weak ($h/\left(k_{\rm B}T_{\rm D}^{0}\right)$=0.1) and strong ($h/\left(k_{\rm B}T_{\rm D}^{0}\right)$=1.0) field amplitudes  are plotted in Fig.~\ref{fig:deltaST3} for three values of the pressure: $p<0$, $p=0$ and $p>0$, i.e., the compressive pressure effect is compared with the stretching effect of a negative pressure. The critical temperatures corresponding to the three considered pressures are shown with dashed vertical lines. It is visible that also for $p\neq 0$ the maxima of $\left|\Delta S_{T,p}\right|$ occur at $T=T_{\rm C}$. The general shape of the thermal dependence of entropy change is conserved, however, the magnitude of $\left|\Delta S_{T,p}\right|$ is pressure-dependent. Namely, $p>0$ reduces the entropy change magnitude, while $p<0$ acts in the opposite direction. Such an effect gives the opportunity to tune the magnetocaloric effect with the external pressure.

The characterization of magnetocaloric effect should be supplemented with the discussion of the second important quantity, namely the adiabatic temperature change. It is interesting to follow the temperature dependence of $|\Delta T_{S,p}|$ for various magnitudes of the magnetic field variation, presented in Fig.~\ref{fig:deltaTS1}. The range of amplitudes is the same as in Fig.~\ref{fig:deltaST2}. In all the cases a single maximum can be observed. First it can be strongly emphasized that the consideration of only purely magnetic component of the entropy within the the Mean Field Approximation leads to the constant entropy for $T>T_{\rm C}$. As a consequence, the adiabatic temperature change for such case would increase unlimitedly for this temperature range, instead of indicating a maximum (such effect was discussed  in our work \cite{szalowski_thermodynamic_201}). However, in the present work, the lattice entropy is included, which shows a non-vanishing temperature dependence in the whole range of temperatures, including $T>T_{\rm C}$. Therefore, the behaviour with a single finite peak seen in Fig.~\ref{fig:deltaTS1} is owing to inclusion of the lattice entropy in the model (however, of course, the temperature change cannot be subdivided into lattice and magnetic subsystem contribution). The location of the maximum is close to $T_{\rm C}$ and when $h$ increases this position tends to shift towards higher temperatures. This contrasts with the behaviour of $|\Delta S_{T,p}|$, which shows the maximum exactly at $T_{\rm C}$ independently on the magnetic field amplitude. On the other hand, the slightly asymmetric shape of the temperature dependence of $|\Delta T_{S,p}|$ follows the trends seen for $|\Delta S_{T,p}|$. The influence of the pressure on adiabatic temperature change can be observed in Fig.~\ref{fig:deltaTS2} both for the positive and negative pressure. Like in the case of isothermal entropy change, the maximum is shifted to higher temperatures by positive pressure. However, the magnitude of $|\Delta T_{S,p}|$ at maximum increases slightly with increasing $p$, which contrasts with the behaviour of $|\Delta S_{T,p}|$ (as its value at maximum was reduced by application of positive pressure).

\section{Final remarks}
\label{sec4}

In our work we have incorporated the static and vibrational lattice properties as well as magnetic subsystem into the thermodynamic description of the model solid, focusing our attention on the influence of the external magnetic field on the system properties. For illustration, we selected a solid with s.c. structure and ferromagnetic exchange interactions between spins $S=1/2$ depending on the lattice deformation, what provided magnetoelastic coupling. In particular, we studied various thermodynamic response functions describing the lattice subsystem (thermal expansion coefficient, compressibility and magnetostriction coefficient) and the effect of the magnetic field on it. We also investigated the behaviour of the entropy and specific heat, discussing the importance of both subsystems. We characterized the magnetocaloric characteristics (isothermal entropy change and adiabatic temperature change) showing the possibility of tuning both quantities with the pressure. The presented results constitute a relatively complete picture of the thermodynamics of the coupled lattice and magnetic system. What is more, they are obtained within a fully self-consistent formalism, based on the construction of the Gibbs energy, with all the thermodynamic relationships fulfilled.

In particular, a new equation for the entropy (Eq.~\ref{eq:Stotal}) has been derived, containing lattice and magnetic contributions, which are interrelated via two equations of state (Eqs.~\ref{eq:state1} and \ref{eq:state2}). It is worth noticing that the lattice entropy has been calculated exactly within Debye model for arbitrary temperature using polylogarithms representation \cite{balcerzak_self-consistent_2017}. On the other hand, the magnetic entropy has been obtained in the Mean Field Approximation, which is the simplest possible, yet useful approach. Further improvement of the magnetic subsystem description can be done using the Pair Approximation method, along the lines presented in Ref.~\cite{szalowski_thermodynamic_2011}.

Possible extensions of the presented approach can involve, for example, inclusion of the electronic degrees of freedom, being a crucial factor for correct description of metals. Also various anisotropies in the behaviour of both elastic properties and of the magnetic interactions can be included in the model. Moreover, the thermodynamics of inhomogeneous systems, like layered or diluted ones, can constitute an interesting field of study. In addition, the magnetics with more complex orderings may exhibit highly interesting behaviour when coupled to the lattice degrees of freedom. Another direction of development may involve multicaloric systems with more order parameters and conjugated fields. 

\section*{Acknowledgments}

This work has been supported under grant VEGA 1/0234/14.

\section*{References}

\bibliographystyle{elsarticle-num}

\end{document}